\documentclass[fleqn,usenatbib,useAMS]{mnras}

\usepackage{longtable}
\usepackage{graphicx}	
\usepackage{amsmath}	
\usepackage{amssymb}	
\usepackage{multicol}        
\usepackage{bm}		
\usepackage{pdflscape}	

\usepackage[T1]{fontenc}
\usepackage{ae,aecompl}
\usepackage{newtxtext,newtxmath}
\usepackage{multirow}

\title{The pride of lions around Messier~105}

\author[Karachentsev et al.]{
Igor D. Karachentsev,$^{1}$\thanks{E-mail: idkarach@gmail.com}
Elena I. Kaisina$^{1}$
and Valentina E. Karachentseva$^{2}$
\\
$^{1}$Special Astrophysical Observatory of the Russian Academy of Sciences, 
Nizhnij Arkhyz, Karachai-Cherkessian Republic, 369167, Russia\\
$^{2}$Main Astronomical Observatory, National Academy of Sciences of Ukraine, Kiev, 03143, Ukraine\\
}

\date{Accepted XXX. Received YYY; in original form ZZZ}

\pubyear{2022}

\begin{document}
\label{firstpage}
\pagerange{\pageref{firstpage}--\pageref{lastpage}}
\maketitle

\begin{abstract}
We undertook a search for new dwarf galaxies in the Leo-I group using the data from the DECaLS digital sky survey. Five new presumed members of this group have been found in a wide vicinity of ${\rm M}\,105 ({\rm NGC}\,3379$). Currently, the group has a population of $83$ galaxies, $33$ of which have measured radial velocities. More than half of the group members belong to early 
types with no signs of ongoing star formation. About a quarter of the 
galaxies are outside the group's virial radius, $R_v = 385$~kpc. The presence of multiple systems with a size of about 15~kpc is evident in the group, but there are no noticeable global flat 
or filamentary substructures. The luminosity function of the group looks to be deficient in galaxies with absolute magnitudes in the interval $M_B = [-18, -15]$ mag. The ${\rm M}\,105$ group is characterized by a 
radial velocity dispersion of $136$~km~s$^{-1}$, orbital mass estimate  $(5.76\pm1.32)\times 10^{12}~M_{\odot}$, and the total mass-to-K-band-luminosity ratio $(17.8\pm4.1) M_\odot/L_\odot$.
The neighboring group of galaxies around ${\rm M}\,66 ({\rm NGC}\,3627$) has a similar virial radius, 
$390$~kpc, velocity dispersion, $135$~km~s$^{-1}$, and total mass-to-luminosity ratio, $(15.6\pm3.9) M_\odot/L_\odot$.  Both groups in the Leo constellation are approaching the Local Group with a velocity of about 100~km~s$^{-1}$. In the background of the ${\rm M}\,105$ group, we noted a group of 6 galaxies with an unusually low virial mass-to-luminosity ratio, $M_T/L_K = (4.1\pm2.2) M_\odot/L_\odot$.

\end{abstract}


\section{Introduction}
The Local Volume (LV), limited by a distance of $D\lesssim11$~Mpc, hosts two dozen galaxies with the luminosity comparable to those of the Milky Way (MW) and the Andromeda galaxy (${\rm M}\,31$). 
Being the dominating dynamic centers of low-population galaxy groups, most of them are surrounded by retinues of dwarf satellites. Measuring the distances and radial velocities of the 
satellites allows one to determine the total mass of the dark halo of the central galaxy. The median value of the total mass-to-$K$-band luminosity ratio for these groups is $M_T/L_K\simeq20 M_\odot/L_\odot$  \citep{karachentsev2021}. In addition to the poor groups, a rich group around the ${\rm M}\,105 ({\rm NGC}\,3379$) galaxy in Leo constellation is located at the far boundary of the LV. It contains 
seven bright galaxies, the luminosity of which are comparable with each other. Early morphological type objects (E, S0), which have depleted their resources for further star formation, 
prevail among them. In terms of population and morphology, the group of galaxies around 
${\rm M}\,105$ appears to be an intermediate link between the poor groups and the closest galaxy clusters Fornax and Virgo.

Searches for new, fainter galaxies in the ${\rm M}\,105$ group (it is also called the Leo-I group) were undertaken repeatedly \citep{ferguson1990,schombert1997,flint2001,karachentsev2004,muller2018}. The fortunate position of the ${\rm M}\,105$ group in the
Arecibo radio telescope survey zone made it possible to measure the radial velocities of dwarf irregular galaxies in the vicinity of ${\rm M}\,105$ 
\citep{stierwalt2009,haynes2011}. The LV galaxy database \citep{kaisina2012} and the Updated Neighboring Galaxy Catalog, UNGC \citep{karachentsev2013} include more than $60$ ${\rm M}\,105$ group candidate members, $20$ of which have measured radial velocities. 
These data were used to estimate the virial mass of the ${\rm M}\,105$ group, which turned out to be in the ($7$--$17$)$\times~10^{12}~M_{\odot}$ range \citep{karachentsev2004,makarov2011,karachentsev2014,karachentsev2015}.

In recent years, the region of the M 105 group has been covered by the DECaLS digital sky survey \citep{dey2019}, which is approximately 1.5 mag deeper than the Sloan Digital Sky Survey \citep{abazajian2009}. This made it possible to check the membership in the group of already known galaxies and to discover new probable members of the group \citep{carlsten2022}. In combination with new measurements of the distances and velocities of galaxies in the group region, this gives reason to return to the estimate of the virial mass of the M 105 group again.

In Section 2, we report the discovery of five new peripheral candidate members of the M 105 group and provide an updated total list of 83 presumable members of the group. In Section 3, we discuss the structural properties of M 105 group. In Section 4, we re-estimate the virial mass of the M 105 group. In Section 5 we discuss structural properties and virial mass estimate for the M 66 group of galaxies neighboring to the M 105 group. Finally, we provide a brief summary of the main results in Section 6.
  
  \section{Search for satellites around M105}
  The paper of \citet{carlsten2022} is an important step in studying the Leo-I galaxy group. Based on the data of the new digital sky survey DECaLS \citep{dey2019} and deep images obtained on large telescopes (CFHT, Subaru, Gemini, Magellan), the authors carried out 
surface ``$g$''- and ``$r$''-band photometry for $60$ presumable group members, estimated the distances to $32$ galaxies using surface brightness fluctuations (SBF), and found $12$ new 
candidate dwarf members of the ${\rm M}\,105$ group. Searches for new low-surface brightness objects were carried out in the virial zone of the group with a radius of about $350$~kpc.

{\centering

\begin{figure*}
\includegraphics[width=2.0\columnwidth]{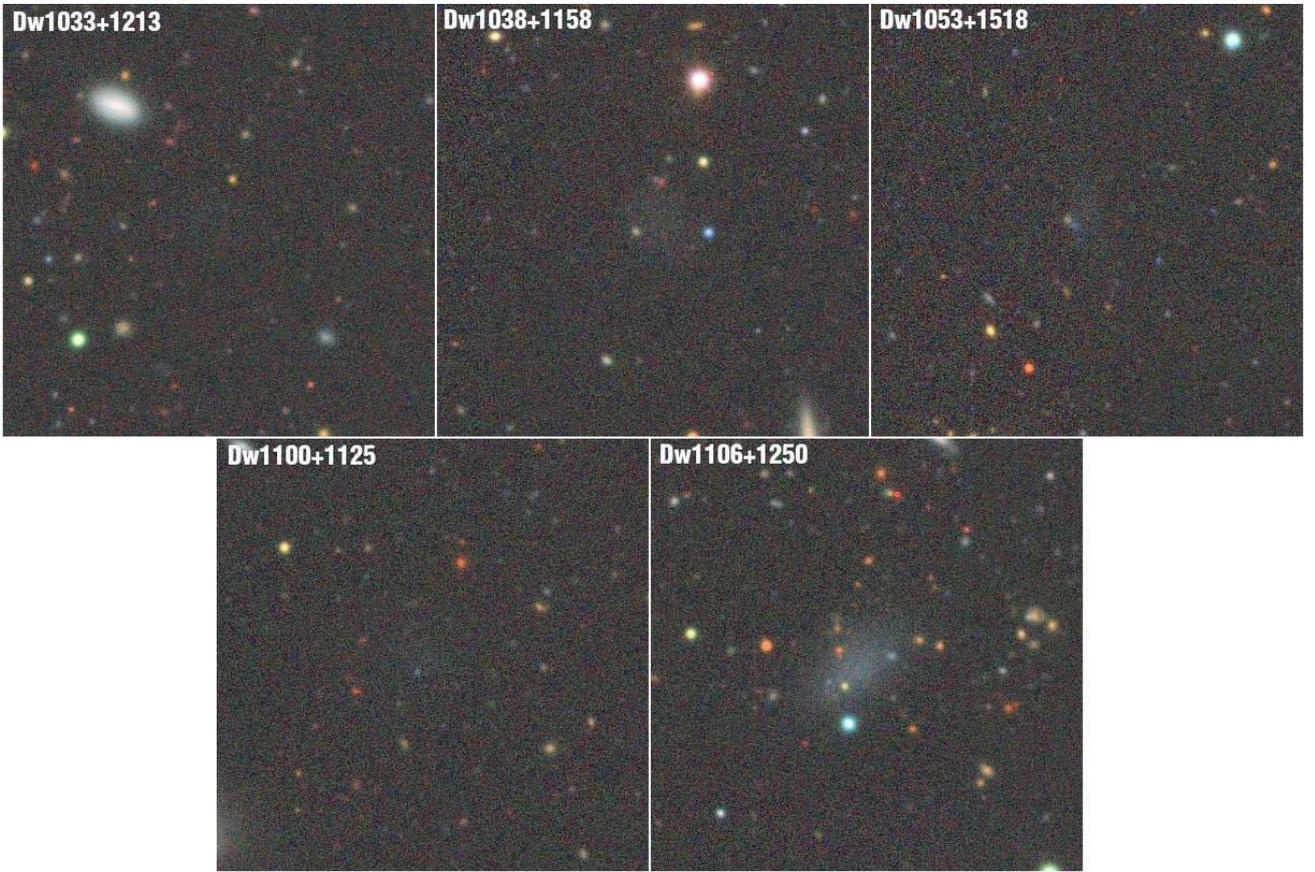}
  \caption{Images of five new presumed ${\rm M}\,105$ satellites from the DECaLS sky survey. The size of the images is $2\arcmin\times 2\arcmin$. North is at the top, east is on the left.}
   \label{figure1}
  \end{figure*}
}
 Evidently, some group members may be located outside the boundaries of its virial radius. We therefore repeated the search for dwarf galaxies in an area approximately four times larger.
 The search for dwarf members of the M 105 group was carried out visually with an emphasis on objects of low surface brightness. The search took into account the shape, structure, and color of galaxy images, which are characteristic of spheroidal and irregular dwarfs. In the virial zone of the group, we independently detected all 12 new dwarf galaxies classified by \citet{carlsten2022} as "confirmed members" based on their distance measurements by the SBF- method. Among the 22 objects classified as "unconfirmed/possible candidate satellites" by \citet{carlsten2022}, we missed 12 cases. All of them turned out to be faint objects of small angular diameters with apparent magnitudes $g\simeq20-21$~mag. We assumed that these 12 objects are rather background galaxies. 
 
 Outside the virial radius of the group, we discovered five new dwarf galaxies: ${\rm Dw}\,1033~+~1213, {\rm Dw}\,1038~+~1158, {\rm Dw}\,1053~+~1518, {\rm Dw}\,1100~+~1125$, and ${\rm Dw}\,1106~+~1250$.  Their DECaLS images are presented in Fig.~1. 
We measured the fluxes of these galaxies in $g$ and $r$ bands using the standard methods of processing in the ESO-MIDAS software package\footnote{\url{https://www.eso.org/sci/software/esomidas//}} for galaxy images from the DESI Legacy Imaging Surveys and obtained the apparent magnitudes according to their description  $m = 22.5-2.5\,log_{10}(flux)$\footnote{\url{https://www.legacysurvey.org/dr10/description/}}. The typical measurment error was about 0.15 mag.
  
A summary list of $83$ presumable ${\rm M}\,105$ group members is presented in Table~\ref{table1}. The columns contain the following: 
name of the galaxy (the new dwarfs detected in this work are highlighted with bold); ${\rm J}2000$ equatorial coordinates in degrees; $B$-band apparent magnitude of the galaxy from the LV galaxies database \footnote{\url{http://www.sao.ru/lv/lvgdb}}(LVGDB); the relation 
$B=g+0.313(g-r)+0.227$, proposed by Lupton 
\footnote{\url{https://www.sdss3.org/dr10/algorithms/sdssUBVRITransform.php\#Lupton2005}}, 
was used for galaxies with measured ``$g$'' and ``$r$'' magnitudes; morphological type: Irr~--- irregular dwarf galaxy, Sph~--- spheroidal dwarf galaxy, Tr~--- intermediate between Irr and Sph,
BCD~--- blue compact dwarf, dE~--- dwarf elliptical; radial velocity relative to the Local Group centroid;  
distance to the galaxy in Mpc; method used to estimate the distance: TRGB~--- by the tip of the red giant branch, Cep~---  by the luminosity of Cepheids, SBF~--- by the 
surface brightness fluctuations, TF~--- by the Tully-Fisher relation between the HI-line width and the galaxy luminosity, mem ~--- by the assumed group membership of the galaxy;  
projected distance of the galaxy from the group center identified with ${\rm M}\,105$, in the assumption that all galaxies have an average distance of $10.8$~Mpc; source of data on the distances of galaxies, reference to which are given at the note of Table 1. 

Data on apparent magnitudes, morphological types, and radial velocities of galaxies are taken from of the LVGDB, version of September 1, 2022. 

It needs to be noted that apparent magnitudes of some faint galaxies in the LVGDB have been estimated roughly by eyeball. We made a comparison of visual B-magnitudes of galaxies from Table 1 with their apparent magnitudes, $g$, measured by \citet{carlsten2020}, based on archival CFHT/MegaCam imaging data. For 30 dwarf galaxies, the average difference in apparent magnitudes is $\langle B - g \rangle = (+0.38\pm0.09)$ mag, which is typical for a mixture of dIrr and dSph galaxies. The root-mean-square magnitude difference, to which the error of our visual estimates makes the main contribution, is $\sigma(B-g)= 0.46$ mag. 

When compiling the list, we 
excluded several galaxies from ${\rm M}\,105$ group members: 
${\rm dw}\,1044+11$ ($104432.7+111610$) as a tidal structure belonging to a distant elliptical galaxy, ${\rm dw}\,1045+14{\rm a}$ 
($104500.9+140619$) as a background galaxy, ${\rm LeG}\,23$ ($105009.1+132901$) as a distant spiral galaxy and ${\rm AGC}\,205268$ 
($104252.4+134428$) whose radial velocity, measured by the HI line, is a result of a mix-up with the HI flux from the neighboring spiral galaxy ${\rm NGC}\,3338$.

In the region of our survey, limited by the coordinates RA = [$10^h30^m$--$11^h10^m$], DEC = [$08\fdg0$--$16\fdg0$], there are 37 other galaxies with radial velocities $V_{\rm LG}<2000$~km~s$^{-1}$. Their list is given in Table~\ref{table2}. Some of these galaxies were previously considered to be members of the Leo-I group. Basic parameters of the galaxies are presented in Table 2 columns, their designations are the same as in
Table~\ref{table1}. The data on apparent magnitudes and radial velocities are taken from NASA/IPAC Extragalactic Database \footnote{\url{http://www.ned.ipac.caltech.edu}} (NED). The sources of data on galaxy distances are shown in Table 2 last column,  reference to which are given at the note of the Table.  
The penultimate column of Table~\ref{table2} gives the $21$-cm line widths, $W_{50}$, at half-maximum intensity level \citep{haynes2011}. As seen, the individual distances to most of the galaxies were 
measured via $W_{50}$ using the Tully-Fisher relation in its usual (TF) or baryonic (bTF) version. The accuracy of the TF method is low for low-luminosity galaxies and gives a distance error of about $30$\%.

\begin{center}     
  \begin{figure}
\includegraphics[width=1.1\columnwidth]{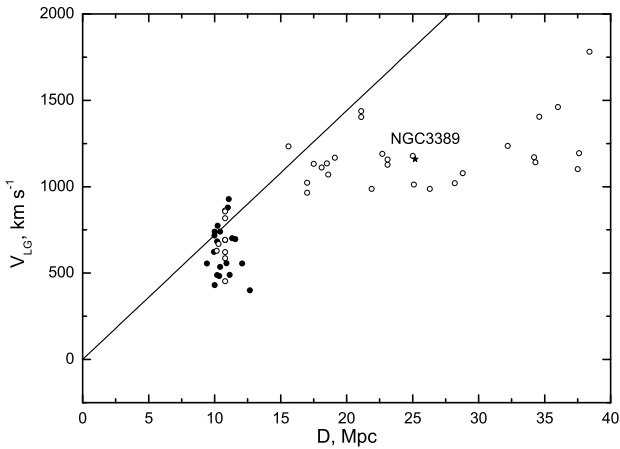}\\[-10mm]
  \caption{Hubble diagram for galaxies in the vicinity of ${\rm M}\,105$. Galaxies with accurate distance estimates (TRGB and SBF) are shown by solid circles. Galaxies with less reliable distance 
estimates (TF and mem) are depicted by open circles. The distant background galaxy
${\rm NGC}\,3389$ with SN measured distance is marked by a star. The straight line corresponds to the unperturbed Hubble flow with the parameter $H_0=72$~km~s$^{-1}$Mpc$^{-1}$.}
  \label{figure2}
  \end{figure} 
   \end{center}

The Hubble ``velocity--distance'' diagram for galaxies in the area under consideration is shown in Fig.~\ref{figure2}. It has a rather specific form, exhibiting a strong nonlinearity. The suggested group members 
around ${\rm M}\,105$ are characterized by an average velocity of
$\langle V_{\rm LG}\rangle=654\pm24$~km~s$^{-1}$ and average distance $10.8$~Mpc. For the Hubble parameter $H_0=72$~km~s$^{-1}$Mpc$^{-1}$ 
\citep{freedman2001}, the ${\rm M}\,105$ group has a negative peculiar velocity of  $-124\pm24$~km~s$^{-1}$ relative to the Local Group centroid. Galaxies with radial velocities $V_{\rm LG}>950$~km~s$^{-1}$ 
from Table~\ref{table2} are scattered across a wide range of distances from $16$~Mpc to $40$~Mpc with a median value of about $25$~Mpc. Almost all of them have high negative peculiar velocities. A significant part of this 
effect is probably due to the TF-distance measurement errors, which scatter the galaxies horizontally by  
$5$--$8$~Mpc. At the same time, the galaxy ${\rm NGC}\,3389$ with an accurate distance estimate of $25.2\pm1.6$~Mpc (marked by a star), measured by supernova luminosity \citep{pejcha2015}, is present among them. It also has a high 
peculiar velocity, $-652\pm115$~km~s$^{-1}$. However, we should note that in an absolute frame of reference associated with the cosmic microwave background (CMB), the radial velocity of ${\rm NGC}\,3389$ amounts to 
$1656$~km~s$^{-1}$, i.e. the galaxy practically rests relative to the CMB reference frame. The kinematic situation in this area of the sky can be presented in such a way so that the complex of distant galaxies with 
$D\gtrsim20$~Mpc rests as a whole with respect to the CMB, while the LV, including the Local Group and the Leo-I group moves in the direction of these distant galaxies with a velocity of about 600~km~s$^{-1}$. 
This velocity, apparently, is due to the motion of the Local Volume directed away from the center of the expanding Local Void, as well as to the collective falling of galaxies towards the Great Attractor \citep{tully2008}.

Although the group members around ${\rm M}\,105$ and the background galaxies do not significantly differ in their radial velocities, their separation by radial distance appears quite obvious.

The distribution of the presumed group members around ${\rm M}\,105$ in equatorial coordinates is presented in Fig.~\ref{figure3}. Dwarf galaxies are shown by circles. Seven of the brighter galaxies with assigned NGC 
numbers are shown by squares. The circle centered on ${\rm M}\,105$ with a radius of $2\fdg04$ (or $385$~kpc for an average $10.8$~Mpc distance to the group) shows the group's virial region. The crosses show the positions 
of $37$ background galaxies from Table.~\ref{table2}.

Among the distant background galaxies, a compact group of six galaxies can be isolated, located to the north-west of ${\rm M}\,105$. These galaxies are highlighted in bold in Table 2 and bolder crosses in Fig.3. The 
Sc galaxy ${\rm NGC}\,3338$ dominates in the group. The group is characterized by an average radial velocity $\langle V_{\rm LG}\rangle=1126$~km~s$^{-1}$, average distance
$19.8$~Mpc,  average projected radius $150$~kpc and an anomalously low radial velocity dispersion $\sigma_V=23$~km~s$^{-1}$. The virial mass of the group is ($2.3\pm1.2)\times11^{10}~M_{\odot}$, and the $K$-band virial 
mass-to-luminosity ratio is equal to only 
$(4.1\pm2.2)M_{\odot}/L_{\odot}$, which indicates a deficit (or total absence) of dark matter in this group.

Note that groups of galaxies with a low $M_T/L_K$ ratio are quite rare. Thus, in the list by \citet{makarov2011} there are 215 groups of galaxies over the whole sky with an average radial velocity $V_{LG} < 3500 $~km~s$^{-1}$ and the number of measured radial velocities $n_v > 5$. Among them, only 15 groups (i.e. $7\%$) have estimates $M_T/L_K < 10 M_\odot/L_\odot$ and 5 groups have $M_T/L_K < 8 M_\odot/L_\odot$. Such "skinny" groups deserve a more detailed study in order to understand the reason for the deficit of dark matter in them.

\section{Structural properties of the group of galaxies around M105}
The presented data allow us to note the following features of the group of galaxies around M 105.
{\centering
\begin{figure*}
\includegraphics[width=1.6\columnwidth]{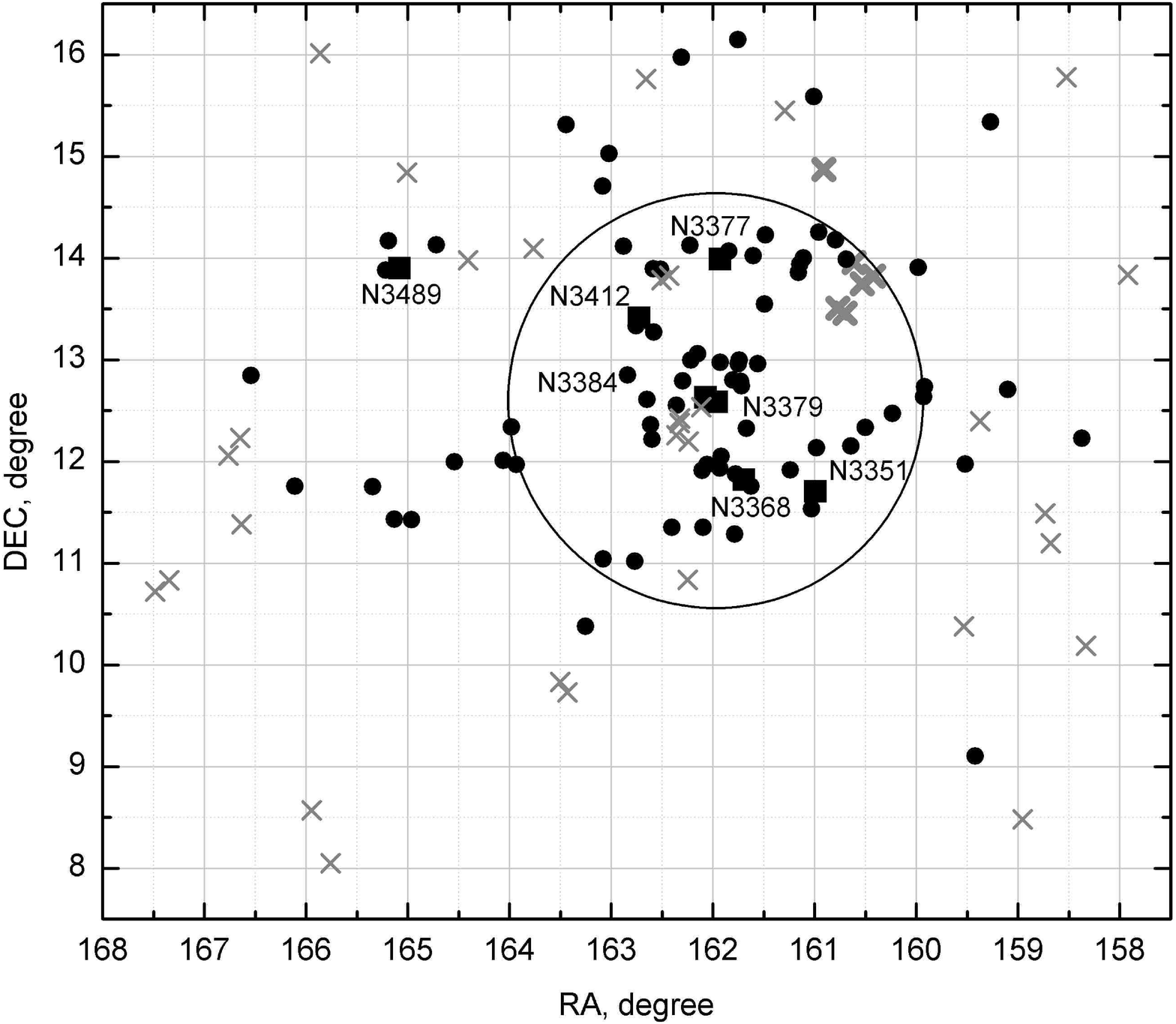}\\[-10mm]
 \caption{Distribution of galaxies in the vicinity of ${\rm M}\,105$ in equatorial coordinates. The suggested ${\rm M}\,105$ group members are shown by circles. Seven of the brightest group members 
with NGC numbers are marked by squares. The crosses show background galaxies with radial velocities less than
$2000$~km~s$^{-1}$. The $385$~kpc ($2\fdg04$) circle corresponds to the group virial radius.}
  \label{figure3}
  \end{figure*}
 } 
 
  About a quarter of the galaxies among the presumable group members ($21$ out of $83$) are located outside the virial radius. Their radial velocities and distances, where available, agree with the membership of these 
galaxies in the group. Therefore, the search area for new group members should not be limited to the region inside the virial radius. The radial velocities and distances of the periphery group members are especially 
important for estimating the total group mass.

  More than half of the galaxies considered ($47$ out of $83$) belong to early morphological types: Sph, E, S0, which have no signs of ongoing star formation. The group exhibits segregation by galaxy type: early-type 
objects have an average projected distance from the center $\langle R_p\rangle=265\pm35$~kpc,  whereas in late-type galaxies (Irr, Tr, S) this quantity amounts to $400\pm45$~kpc. 
 Nonetheless, some early-type galaxies (${\rm NGC}\,3489, {\rm Dw}\,1033+1233$) meet at distances $R_p>600$~kpc.
 
  It is of interest to note that the ${\rm M}\,105$ group does not have any blue compact dwarfs (BCD), even though such objects, rich in gas, could be easily detected in the ALFALFA survey \footnote{\url{http://egg.astro.cornell.edu/}}. In some groups, for example, 
around ${\rm M}\,81$, BCD galaxies form a noticeable part of the population.

\begin{figure}
\vspace*{10mm}
\begin{tabular}{c}\\[-20mm]
\includegraphics[width=0.5\textwidth,clip]{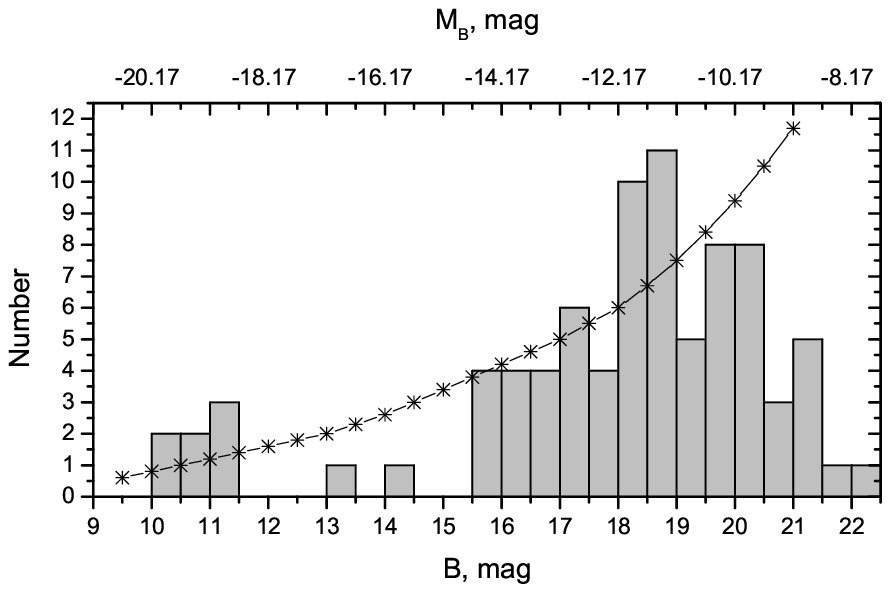}\\[-10mm]
\includegraphics[width=0.5\textwidth,clip]{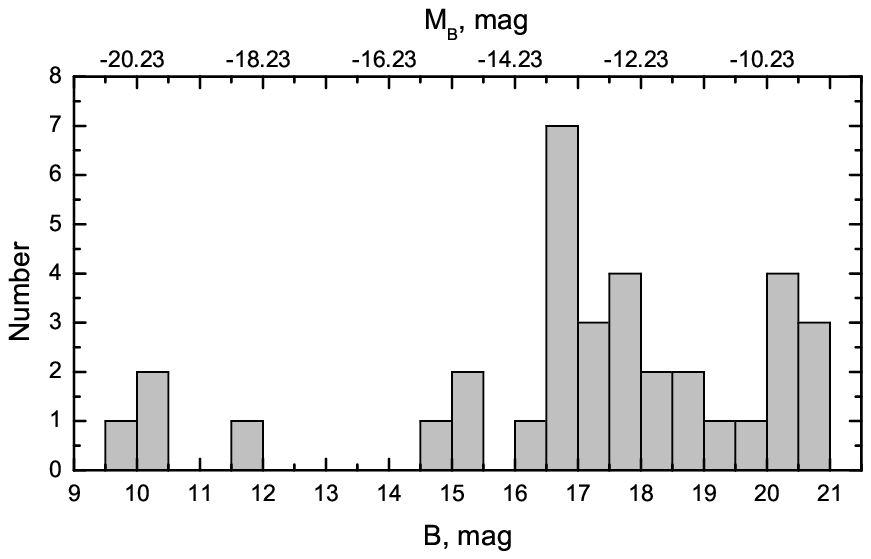}\\[-5mm]
\end{tabular}
 \caption{Luminosity function for the suggested members of M105 (upper panel) and M66 (lower panel).
 The solid line on the upper panel indicates the luminosity function for the Local volume population.}
  \label{figure4}
  \end{figure}

  The distribution of the group galaxies by apparent and absolute $B$-magnitude is presented in the upper panel of Fig.~\ref{figure4}. The solid line shows the luminosity function of the LV galaxies according to \citet{driver2003}, which is well described by the standard \citet{schechter1976} function
 \begin{equation}    
    F(M) = c [10^{0.4(M^*-M)}]^{1+\alpha}\times exp[-10^{0.4(M^*-M)}]
 \end{equation} 
 with parameters $\alpha$ = -1.2 and $M^*=-21.5$ mag. In the range of $M_B < -11.0$ mag, where the incompleteness of the sample is not yet affected, the probability of agreement between the observed and expected distributions is $p=0.08$ according to the $\chi^2$ test.
 The observed distribution shows a marked deficit of intermediate luminosity galaxies (similar to Magellanic Clouds) with absolute magnitudes [$-18^m, -15^m$]. Such a gap may be related to the peculiarities of the formation of this group or the galaxy merging process in it.

  Flat or filamentary substructures were detected in the nearby groups around the Milky Way, ${\rm M}\,31$, Centaurus-A galaxies, consisting of dwarf satellites \citep{pawlowski2013,ibata2014,libeskind2015}.  The panorama of the Leo-I group (Fig.~\ref{figure3}) exhibits a chain of eight galaxies, located across the diagonal in the NW direction from the ${\rm NGC}\,3351$ ($161\fdg0+11\fdg7)$ spiral. 
However, the radial velocities known for five of them do not show any specific alignment that could indicate a physical nature of this filament. The group exhibits also several dwarf pairs and triplets with mutual projected 
separations of the components $3\arcmin$--$5\arcmin$, i.e. $10$--$15$~kpc. Some of them have close radial velocities (${\rm KK}\,94$ and ${\rm LeG}\,21$) or a mutual morphological similarity (${\rm dw}\,1052+1352$ and 
${\rm dw}\,1050+1353$).  Such substructures deserve a more detailed investigation.

 \citet{schneider1989} studied the central region of this group in the HI line and discovered nine intergalactic hydrogen clouds, which form a giant ring-like structure with a diameter of about $1\degr$ or approximately $200$ kpc.
The radial velocities of the HI clouds are in the $V_{\rm LG}=600$--$900$~km~s$^{-1}$ range, which implies their Leo-I group membership. \citet{mihos2018} obtained deep frames of one HI cloud near a Sab galaxy 
${\rm NGC}\,3368$ and detected a diffuse blue object of very low surface brightness (about $29^m$/sq.arcsec). This object, 
${\rm BST}\,1047+1156$, is also noticeable in the ultraviolet images, taken with GALEX space telescope. The unusual dwarf system with a hydrogen mass of about $4.5\times 10^7~M_{\odot}$ has a low star formation rate of about 
$5\times 10^{-5}~M_{\odot}$/year. About $99$\% of its mass is in gaseous state.

\citet{sameer2022} studied spectra of $11$ quasars located behind the ${\rm M}\,105$ group in a $7\times 6$ square degree area. From the absorption lines corresponding to the radial velocity range of the Leo-I group, 
they determined that the intergalactic medium of the group has a high metallicity of the order of or higher than the solar metallicity. Based on this, the authors conclude that the huge ring-like structure made of HI clouds 
does not have the relic origin but was formed as a result of tidal interaction of the brightest group members. This result, as well as the presence of morphological segregation of galaxies, indicates that the Leo-I group is at a 
rather advanced stage of its dynamic evolution.

\section{Virial mass of the M105 group}
Several methods of estimating the total mass of a galaxy group can be encountered in the literature. They are all based on the virial balance of kinetic and gravitational energy of the system. The differences between them are 
associated with different assumptions about the character of galaxy motion and different considerations of the projection factors.

\citet{limber1960} proposed estimating the virial mass of a group consisting of $n$ galaxies as

\begin{equation}
 M_v= (3\pi/2)\times n\times(n-1)^{-1}\times G^{-1}\times \sigma_V^2\times R_h,
 \end{equation}
 
 where $\sigma_V$ is the galaxy radial velocity dispersion relative to the center, $R_h$ is the average harmonic distance between the galaxies in sky projection, and $G$ is the gravitational constant. For the group around 
${\rm M}\,105$ we obtained  $\sigma_v=136$~km~s$^{-1}$ and $R_h=191$~kpc from  
$33$ galaxies with measured radial velocities, which gives a group virial mass estimate of $M_V=3.88\times10^{12}~M_{\odot}$. \citet{heisler1985} performed numerical n-body simulations of groups of galaxies with equal masses and found that for groups with $n\sim10$, the virial mass estimator gives a typical error of mass about $10^{0.15}$.


According to \citet{tully2015}, the total virial mass of a rich group may be determined as
\begin{equation}
M_T=(\beta\times\pi/2)\times G^{-1}\times\sigma_V^2\times\langle R_p\rangle,
\end{equation}
where $\langle R_p\rangle$ is the average projected group radius, and $\beta\simeq2.5\pm0.1$ addresses the transition from radial velocity dispersion to volumetric dispersion in the assumption of a weak anisotropy. 
For $\sigma_V=136$~km~s$^{-1}$ and $\langle R_p\rangle$=$318\pm35$~kpc we derive a mass estimate of $ M_T=$$(5.35\pm1.25)$$\times 10^{12}~M_{\odot}.$

Another ``orbital'' mass estimation method \citep{bahcall1981}, \citep{karachentsev2014} assumes that small group members move around the massive center by arbitrarily oriented elliptical orbits with an average eccentricity of 
$\langle e^2\rangle=1/2$ \citep{barber2014}. In this case,
\begin{equation}
M_{\rm orb}=(16/\pi)\times G^{-1}\times \langle \Delta V^2\times R_p\rangle.
\end{equation}

Based on $33$ galaxies with known radial velocities we get $M_{\rm orb}=(5.76\pm1.32)\times 10^{12}~M_{\odot}$. 

Note that in the case of a random Kepler orbit orientation with an eccentricity of $e^2=1/2$,  the relative 
$M_{\rm orb}$ estimate variation due to the projection factors amounts to
\begin{equation}
\sigma(M_{\rm orb})/\langle M_{\rm orb}\rangle=[(256-10\pi^2)/10\pi^2]^{1/2}=1.2625.
\end{equation}
This value is in wonderful agreement with the relative variation of $M_{\rm orb}$ obtained by the scatter of individual 
$M_{\rm orb}$ estimates made for different group members: $\sigma(M_{\rm orb})=7.48\times 10^{12}~M_{\odot}$ and $\sigma(M_{\rm orb})/M_{\rm orb}=1.2986$.

According to the LV galaxies database, the total $K_s$-band luminosity of the ${\rm M}\,105$ group galaxies is $L_K=3.23\times10^{11}\,L_{\odot}$. Adopting the obtained $M_{\rm orb}$ values 
as an optimal virial mass estimate of the group, we have the total mass-to-luminosity ratio $M_{\rm orb}/L_K=(17.8\pm4.1)~M_{\odot}/L_{\odot}$.

Tully (2015) found that the group virial mass and virial radius are connected through an empirical relation
\begin{equation}
(R_V/215)=(M_V/10^{12})^{1/3},
\end{equation}
 where the radius $R_V$ is expressed in kpc, and the mass is in solar units. For the derived orbital mass estimate, the virial radius is $R_V= 385$~kpc 
or $2\fdg04$ (shown by the circle in Fig.~\ref{figure3}).

{\centering
    \begin{figure*}
\includegraphics[width=1.3\columnwidth]{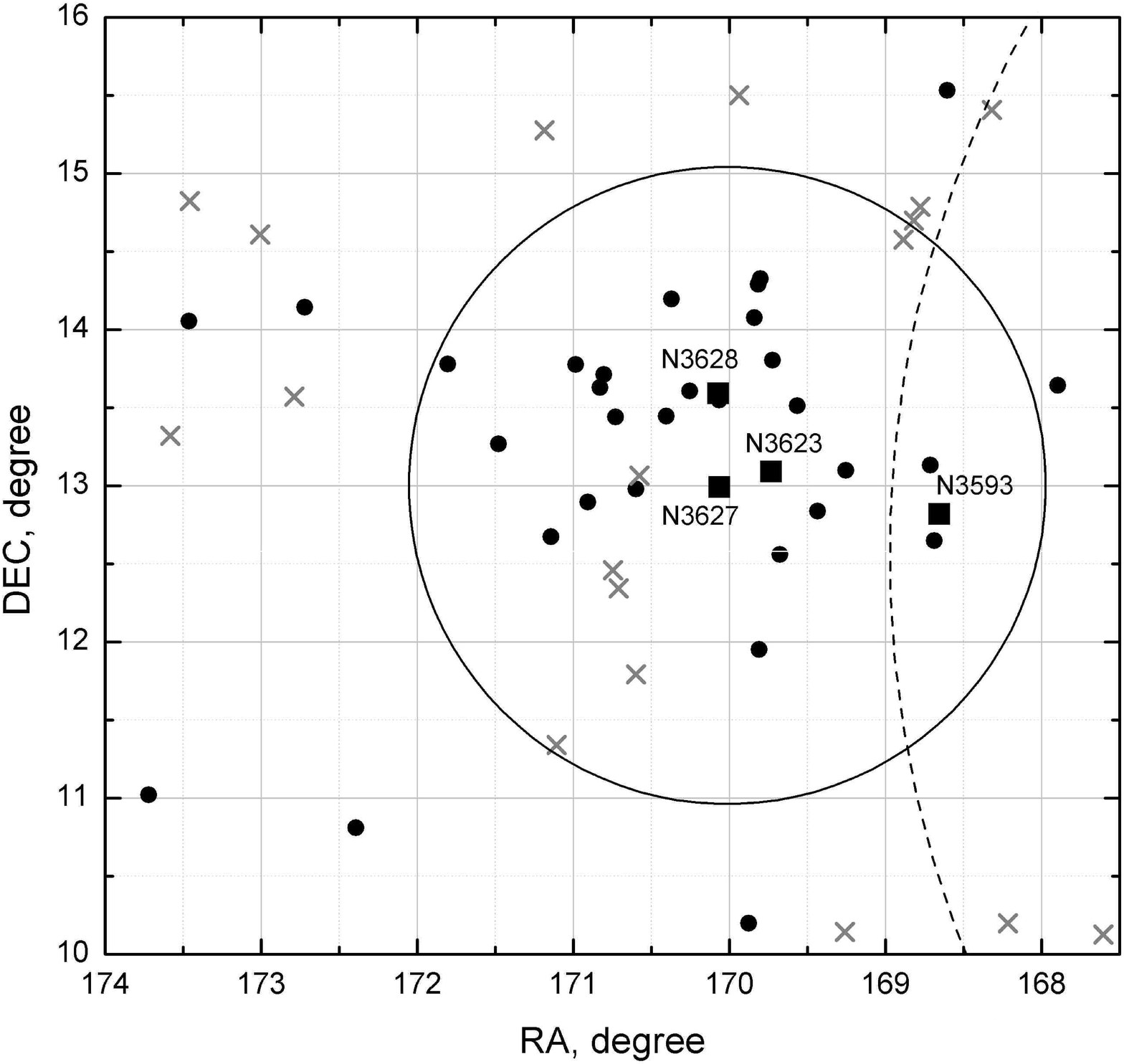}
 \caption{Distribution of galaxies in the vicinity of ${\rm M}\,66$. The symbols for the suggested group members and background galaxies are the same as in Fig.~\ref{figure3}. The circle corresponds 
to the virial radius $390$~kpc ($2\fdg01$). The dashed arc on the right shows the zero velocity sphere radius for the neighboring ${\rm M}\,105$ group.}
 \label{figure5}
  \end{figure*}
} 
 
 \section{Neighboring group of galaxies around M66}
 Considering the distribution of galaxies in the local universe, \citet{tully1988} noted that the galaxy group associated with ${\rm M}\,105$ is part of the diffuse extended structure named Leo~Spur. This cloud also hosts 
close groups around galaxies 
${\rm NGC}\,3521$ ($D=10.7$~Mpc) and ${\rm NGC}\,3115$ ($D=10.2$~Mpc), the structures and populations of which are described in \citet{karachentsevmak2022}. These groups are separated from ${\rm M}\,105$ by angular 
distances $13\degr$ and $23\degr$, correspondingly. The closest neighbor of ${\rm M}\,105$ is the group of galaxies around ${\rm M}\,66 ( {\rm NGC}\,3627$), the distance to which is $11.12\pm0.56$~Mpc \citep{hoyt2019}. 
The angular distance between the centers of these groups is equal to $7\fdg9$.
 
The group ${\rm M}\,66$, as is the case for group ${\rm M}\,105$, is located in the ALFALFA HI survey zone \citep{haynes2011}, which contributed to measuring the radial velocities for many irregular dwarf members of 
the group. Searches for dwarf galaxies around ${\rm M}\,66$ were undertaken by various authors \citep{muller2018,carlsten2022,karachentsev2022}. To date, 35 ${\rm M}\,66$ candidate members, 
$22$ of which have measured radial velocities, were detected in the RA = $167\fdg5$--$174\fdg0$, DEC = $10\fdg0$--$16\fdg0$ region. 

 Data on the presumable ${\rm M}\,66$ group members are given in Table~\ref{table3}. Its structure and designations of the presented parameters are the same as in Table~\ref{table1}. Data on apparent magnitudes, morphological types, and radial velocities of galaxies are taken from the latest version of the LV database.The sources of data on galaxy distances are shown in the Table last column, reference to which are given at the note of the Table.
 
 The distribution of galaxies in the area recently searched by us \citep{karachentsev2022} for new group members is shown in Fig.~\ref{figure5}. Four bright galaxies having NGC numbers are marked by squares, the rest of the dwarf galaxies are shown by circles. The fact that this group has been 
studied in much less detail than ${\rm M}\,105$ comes to attention. Only one galaxy, ${\rm M}\,66$ itself, has an accurate distance estimate obtained by the TRGB method (there are $15$ such galaxies in the 
${\rm M}\,105$ group), although both groups are located practically at the same distance from the observer. A deficit of intermediate luminosity galaxies like the Magellanic Clouds, with the absolute magnitude interval $M_B= [-16^m, -18^m]$, is evident 
in the luminosity function for group ${\rm M}\,66$ (see lower panel in Fig.~\ref{figure4}). Early type galaxies (Sph, dE, S0) make up $40\%$ of the total population. Unlike ${\rm M}\,105$, the ${\rm M}\,66$ group has four 
blue compact dwarfs (BCDs).

 In the considered region, $18$ background galaxies are known, with radial velocities $V_{\rm LG}<2000$~km~s$^{-1}$. These galaxies are marked by crosses in Fig.~\ref{figure5}. Distance estimates were obtained for $13$ 
of them based on the Tully-Fisher relation \citep{karachentsev2015}. Given the large scatter of distance estimates for the background galaxies, their median value is equal to $22$~Mpc. Comparing it to the median radial 
velocities $\langle V_{\rm LG}\rangle =1150$~km~s$^{-1}$ 
and $\langle V_{\rm CMD}\rangle =1590$~km~s$^{-1}$, we conclude that the conglomerate of galaxies behind the ${\rm M}\,66$ group approaches the Local Volume with a typical velocity of about $500$~km~s$^{-1}$, but rests relative 
to the cosmic microwave background. A similar situation also takes place for the galaxies behind the ${\rm M}\,105$ group, as was noted above.
 
According to the data in Table~\ref{table3}, the ${\rm M}\,66$ group is characterized by the following parameters: average radial velocity 
$\langle V_{\rm LG}\rangle~=~646$~km~s$^{-1}$, radial velocity dispersion $\sigma_V=135$~km~s$^{-1}$, average projected separation of the members $\langle R_p\rangle=270$~kpc, and the total orbital mass estimate 
$M_{\rm orb}=(5.94\pm1.50)\times 10^{12}~M_{\odot}$. For the total group luminosity $L_K=3.83\times 10^{11}~L_{\odot}$, the total mass-to-luminosity ratio is $M_T/L_K=(15.6\pm3.9)~M_{\odot}/L_{\odot}$, which coincides 
within the margin of error with the same ratio for group ${\rm M}\,105$. The virial radius of the ${\rm M}\,66$ group determined from expression ($6$) is $R_V=390$~kpc or $2\fdg01$. It is shown in Fig.~\ref{figure5} by a circle centered on ${\rm M}\,66$.

Besides the virial radius, the zero velocity sphere radius $R_0$, where the dynamic influence of the group dominates over general cosmological expansion, is an important group parameter. For a typical relation 
$R_0\simeq 3.5\times R_V$, the radius $R_0$ for both neighboring groups is $7\fdg1$. A fragment of an arc of radius $R_0$ around the ${\rm M}\,105$ group center is shown in Fig.~\ref{figure5} in dashes. The mutual distance 
between the centers of these groups,  $7\fdg9$, turns out to be somewhat larger than their $R_0$ radii. Therefore, future merging of both groups into a single system appears unlikely.

We should note that the ${\rm M}\,66$ group, as is the case with ${\rm M}\,105$, shows signs of active galaxy interaction. In the ${\rm M}\,66$ group, the second brightest spiral galaxy ${\rm NGC}\,3628$ exhibits strong 
distortions at the periphery. A tidal tail of very low surface brightness stretches to the east of ${\rm NGC}\,3628$ \citep{jennings2015}, the length of which reaches $42\arcmin$  or $135$~kpc. This structure is well seen
also in the HI line \citep{stierwalt2009}. The dwarf galaxy 
${\rm dw}\,1120+1337={\rm NGC}\,3628,{\rm UCD}\,1$ is the brightest part of this tail and was probably destroyed by the tidal influence of ${\rm NGC}\,3628$. The far end of the eastern tail hosts an irregular dwarf galaxy 
with low surface brightness ${\rm AGC}\,215414$, the hydrogen mass of which is $25$ times higher than its stellar mass \citep{nikiel2014}.

The main parameters of the ${\rm M}\,66$ group in comparison with those for the ${\rm M}\,105$ group are given in Table~\ref{table4}.

\section{Concluding remarks}
The group of galaxies around ${\rm M}\,105$ is one of the richest groups in the Local Volume. Unlike most other LV groups, not one but several galaxies dominate in it with a luminosity comparable to that of 
the Milky Way. The group is conveniently located in the ``blind'' HI strip survey carried out with the Arecibo radio telescope, which allowed radial velocities to be measured for many dwarf members of this system. In recent 
years the population of the group has significantly increased due to searches for new dwarf galaxies in the images of the DECaLS multi color digital sky survey. In addition to the searches for new objects in the group virial 
zone \citep{carlsten2022}, we extended the search area to distant outskirts of the group and discovered another five candidate group members.

 Currently the Leo-I group includes $83$ presumable members, $33$ of which have measured radial velocities. Using the projected separations and radial velocities of these galaxies, we determined their total group mass using 
four different methods. The estimates turned out to be in the
$(3.9$--$5.8)\times 10^{12}~M_{\odot}$ range. Early type galaxies without signs of star formation exhibit a known effect of stronger concentration towards the group center. The presence of small multiple systems is noticeable 
in Leo-I, but no large co-planar structure has been detected.

Another group of galaxies around ${\rm M}\,66$ neighbors the ${\rm M}\,105$ group. Both the groups are separated from the observer by similar distances, $10.8\pm0.5$~Mpc and $11.1\pm0.5$~Mpc, 
correspondingly. The similarity of groups
$ {\rm M}\,105$ and ${\rm M}\,66$ also manifests itself in other parameters: the average radial velocity, the radial velocity dispersion, the average projected radius, the total luminosity, and the orbital mass estimate. 
Both the groups have a luminosity function gap in the absolute magnitude interval [$-15^m, -18^m$], corresponding to the luminosity of the Magellanic Clouds. At the same time, the number
population of the ${\rm M}\,66$ group turned out to be twice smaller than that of ${\rm M}\,105$, which is in no way due to observational selection.

In the background of the ${\rm M}\,105$ group, we noted a group of 6 galaxies with an unusually low virial mass-to-luminosity ratio, $M_T/L_K = (4.1\pm2.2) M_\odot/L_\odot$, which indicates a deficit or total absence of dark matter in this group. 

The conglomerate of distant background galaxies with $V_{\rm LG}~=~1000$--$2000$~km~s$^{-1}$ radial velocities behind both groups has a median distance of $D\sim25$~Mpc and $22$~Mpc, correspondingly. The collection of distant galaxies behind the group is in a state of rest relative to the cosmic microwave background, and both the  ${\rm M}\,105$ and ${\rm M}\,66$ groups move 
towards the distant background with a characteristic peculiar velocity of about 
$600$~km~s$^{-1}$.
 
\section*{Acknowledgements}
We are grateful to the anonymous referee for his/her constructive comments that helped to improve the paper.
 This work has made use of the DECaLS sky survey, The Arecibo Legacy Fast ALFA (ALFALFA), the
NASA/IPAC Extragalactic Database, and the revised version of the Local Volume galaxy database. The Local Volume galaxies database has been updated within the framework of grant $075-15-2022-262$ ($13.{\rm MNPMU}.21.0003$) of the Ministry of Science and Higher Education of the Russian Federation.

\section*{Conflict of Interest}
The authors declare no conflict of interest.

\section*{Data Availability}
The data  underlying this article are available in the Local Volume galaxy database https://www.sao.ru/lv/lvgdb/. The additional data will be shared on reasonable request to the corresponding author.

\label{lastpage}
\clearpage
\onecolumn
\begin{longtable}{lcclcrllc}
\caption{Suggested members of the ${\rm M}\,105$ group.} 
\label{table1}  \\
\hline\hline 
\multicolumn{1}{c}{Name}&
\multicolumn{1}{c}{RA J2000.0 Dec.} & 
\multicolumn{1}{c}{$B_t$} & 
\multicolumn{1}{c}{Type}&
\multicolumn{1}{c}{$V_{\rm LG}$} & 
\multicolumn{1}{c}{$D$}& 
\multicolumn{1}{c}{Method} & 
\multicolumn{1}{c}{$R_p$}  &
\multicolumn{1}{c}{Refs}\\
 & deg & mag &  & km~s$^{-1}$ & Mpc & & kpc &  \\ 
\hline 
\endfirsthead

{{\tablename\ \thetable{} -- continued}} \\
\hline\hline
\multicolumn{1}{c}{Name}&
\multicolumn{1}{c}{RA J2000.0 Dec.} & 
\multicolumn{1}{c}{$B_t$} & 
\multicolumn{1}{c}{Type}&
\multicolumn{1}{c}{$V_{\rm LG}$} & 
\multicolumn{1}{c}{$D$}& 
\multicolumn{1}{c}{Method} & 
\multicolumn{1}{c}{$R_p$}  &
\multicolumn{1}{c}{Refs}\\
 & deg & mag &  & km~s$^{-1}$ & Mpc & & kpc &  \\ 
\hline
\endhead
\hline 
\endfoot
\endlastfoot    
{\bf 
Dw1033+1213}   &  158.372 +12.229 &  22.1 &  Sph         &   -    &  10.8   &   mem        & 662 &   \\
NGC3299	       &  159.099 +12.707 &  13.3 &  Sdm   	 &  453   &  10.8   &   mem        & 525 &    \\ 
AGC205165      &  159.270 +15.337 &  16.4 &  Irr 	 &  586   &  10.8   &   mem        & 716 &    \\
dw1037+09      &  159.420 +09.105 &  17.9 &  Irr  	 &   -    &  10.8   &   mem        & 804 &    \\
{\bf 											   	     
Dw1038+1158}   &  159.520 +11.973 &  20.5 &  Sph         &   -    &  10.8   &   mem        & 462 &   \\
LeG04	       &  159.917 +12.734 &  18.7 &  Tr  	 &   -    &  10.8   &   mem        & 376 &    \\
LeG05	       &  159.929 +12.635 &  16.8 &  Tr  	 &  629   &  10.8   &   mem        & 370 &    \\
LeG06	       & 159.982 +13.908  & 18.3  & Irr 	 &  863   &  10.8   &   mem        & 440 &    \\
UGC05812       & 160.236 +12.473  & 15.5  & Im  	 &  857   &  10.8   &   mem        & 314 &    \\
FS04	       & 160.502 +12.335  & 15.7  & Im  	 &  621   &  10.8    &  mem        & 269 &    \\
LeG09	       & 160.643 +12.151  & 17.4  & Sph 	 &   -    &  10.19   &  TRGB   & 253 &  (1)  \\
dw1042+1359    & 160.689 +13.990  & 21.2  & Sph 	 &   -    &  10.51   &  SBF    & 350 &  (2)   \\
dw1043+1410    & 160.795 +14.180  & 20.1  & Sph 	 &   -    &  11.92   &  SBF    & 366 &  (2)   \\
dw1043+1415    & 160.957 +14.255  & 20.2  & Sph 	 &   -    &  12.24   &  SBF    & 362 &  (2)   \\
LeG10	       & 160.979 +12.133  & 19.2  & Irr 	 &   -    &  10.8    &  mem    & 197 &        \\
NGC3351 (M95)  & 160.990 +11.704  & 10.6  & Sb           &  623   &  9.95   &   TRGB   & 241 &  (3) \\
LeG11          & 161.008 +15.589  & 18.8  & Tr  	 &   -    &  10.8   &   mem    & 593 &        \\
LeG12	       & 161.031 +11.534  & 18.5  & Irr 	 &   -    &  10.8   &   mem    & 259 &        \\
dw1044+1359    & 161.108 +13.999  & 21.0  & Sph 	 &   -    &  10.60   &  SBF    & 306 &  (2)   \\
AGC205445      & 161.147 +13.940  & 16.4  & Tr  	 &  490   &  11.13   &  SBF    & 293 &  (2)   \\
dw1044+1351    & 161.159 +13.859  & 19.9  & Sph 	 &   -    &  11.49   &  SBF    & 279 &  (2)   \\
LeG13	       & 161.239 +11.916  & 17.7  & Irr 	 &  718   &   9.99   &  SBF    & 180 &  (2)   \\
dw1045+14b     & 161.484 +14.227  & 20.0  & Sph 	 &   -    &  10.8    &  mem    & 319 &        \\
dw1045+13      & 161.492 +13.548  & 19.8  & Sph 	 &   -    &  10.8    &  mem    & 199 &        \\
LeG14	       & 161.559 +12.961  & 18.7  & Sph 	 &  740   &  10.00   &  TRGB   & 101 &  (1)  \\
KK93	       & 161.603 +14.024  & 18.0  & Sph 	 &   -    &   9.51   &  TRGB   & 277 &  (1)  \\
LeG16	       & 161.626 +11.756  & 18.4  & Sph 	 &   -    &  11.02   &  SBF    & 166 &  (2)   \\
LeG17	       & 161.672 +12.327  & 17.0  & Sph 	 &  880   &  10.99   &  SBF    & 70  &  (2)  \\
NGC3368 (M96)  & 161.691 +11.820  & 10.1  & Sab   	 &  740   &  10.42   &  Cep    & 150 &  (4)   \\
LeG18	       & 161.720 +12.742  & 18.2  & Irr  	 &  488   &  10.19   &  TRGB   & 52  &  (1) \\
LeG19	       & 161.728 +12.789  & 18.8  & Sph 	 &   -    &  11.68   &  SBF    & 57  &  (2)  \\
KK94	       & 161.739 +12.998  & 17.5  & Tr  	 &  684   &  10.19   &  TRGB   & 87  &  (1) \\
dw1047+16      & 161.754 +16.146  & 18.3  & Sph 	 &   -    &  10.8    &  mem    & 673 &        \\
LeG21	       & 161.752 +12.960 &  19.2 &  Irr          &  696   &  11.56   &  SBF    & 80  &  (2)  \\
dw1047+1153    & 161.775 +11.878 &  21.4 &  Sph          &  -     &   9.35   &  SBF    & 136 &  (2)  \\
N3368dwTBG     & 161.784 +11.285 &  19.6 &  Sph          &  -     &  10.8    &  mem    & 244 &        \\
M96-DF7	       & 161.805 +12.802 &  18.9 &  Sph          &  -     &  10.19   &  TRGB   &  50 &  (1)  \\ 
DDO088	       & 161.843 +14.069 &  14.4 &  Sm           &  431   &  10.00   &  TRGB   & 281 &  (5) \\
M96-DF2	       & 161.919 +12.049 &  20.0 &  Sph          &  -     &  10.62   &  TRGB   & 100 &  (1) \\
dw1047+1258    & 161.926 +12.975 &  20.0 &  Sph          &  -     &  11.00   &  SBF    &  74 &  (2)  \\ 
NGC3377	       & 161.926 +13.986  & 11.2  & E     	 & 536    &  10.42   &  TRGB   & 262 &  (3)  \\
BST1047+1156   & 161.932 +11.934  & 20.0  & Irr 	 & 818    &  10.8    &  mem    & 122 &        \\
NGC3379 (M105) & 161.957 +12.582  & 10.2  & E     	 & 774    &  10.23   &  TRGB   &  0  &  (6) \\
M96-DF1        & 162.055 +11.968  & 19.8  & Sph 	 &  -	  &   10.42  &  TRGB   & 116 &  (1) \\ 
NGC3384	       & 162.070 +12.629  & 10.9  & S0    	 & 556    &   9.42   &  TRGB   & 23  &  (3) \\
dw1048+1121    & 162.096 +11.352  & 20.5  & Sph 	 &  -	  &   10.8   &  mem    & 231 &        \\
dw1048+1154    & 162.104 +11.912  & 19.7  & Sph 	 &  -	  &   10.75  &	SBF    & 128 &  (2)  \\ 
dw1048+13      & 162.149 +13.060  & 20.1  & Sph 	 &  -	  &   10.62  &	TRGB   &  96 &  (1) \\
dw1048+1259    & 162.217 +12.995  & 20.3  & Sph 	 &  -	  &   11.81  &	SBF    &  91 &  (2)  \\
AGC200596      & 162.224 +14.124  & 15.7  & dE  	 & 400    &   12.67  &  SBF    & 292 &  (2)    \\
dw1049+12a     & 162.298 +12.792  & 19.8  & Tr  	 &  -	  &   11.91  &	SBF    &  74 &  (2)  \\
dw1049+15      & 162.310 +15.973  & 18.8  & Tr  	 &  -	  &   10.8   &  mem    & 642 &         \\
dw1049+12b     & 162.358 +12.552  & 19.3  & Sph 	 &  -	  &   10.81  &	TRGB   &  73 &  (1) \\
LeG22	       & 162.404 +11.351  & 18.0  & Tr 	         &  -	  &   10.8   &  mem    & 244 &         \\
dw1050+1352    & 162.515 +13.889  & 19.6  & Sph 	 &  -	  &   10.8   &  mem    & 264 &         \\
UGC05944       & 162.579 +13.272  & 15.6  & dE  	 & 928    &   11.07  &  SBF    & 172 &  (2)    \\
dw1050+1353    & 162.585 +13.896  & 19.2  & Sph 	 &  -	  &   10.8   &  mem    & 271 &         \\
dw1050+1213    & 162.600 +12.218  & 21.3  & Sph 	 &  -	  &   11.19  &	SBF    & 136 &  (2)  \\
KK96	       & 162.613 +12.360  & 17.3  & Sph 	 &  -	  &   10.00  &	TRGB   & 127 &  (1) \\
dw1050+1236    & 162.649 +12.609  & 20.6  & Tr 	         &  -	  &   11.25  &	SBF    & 126 &  (2)  \\
NGC3412        & 162.722 +13.412  & 11.4  & S0    	 & 702    &   11.32  &  SBF    & 208 &  (7)    \\
LSBC D640-11   & 162.756 +13.334  & 16.7  & Sph 	 & 557    &   10.89  &  SBF    & 202 &  (2)    \\
dw1051+11      & 162.766 +11.020  & 18.1  & Sph 	 &  -	  &   10.8   &  mem    & 327 &         \\
LeG26	       & 162.838 +12.849  & 16.6  & Sph 	 & 483    &   10.34  &  SBF    & 168 &  (2)    \\
AGC205540      & 162.881 +14.115  & 18.0  & Tr  	 & 691    &   10.8   &  mem    & 332 &         \\
AGC205544      & 163.020 +15.030  & 17.1  & dE  	 & 692    &   10.8   &  mem    & 501 &         \\
AGC202456      & 163.081 +11.043  & 16.2  & dE  	 & 669    &   10.8   &  mem    & 353 &         \\
LeG27	       & 163.082 +14.708  & 18.6  & Tr  	 &  -	  &   10.8   &  mem    & 451 &         \\
LeG28	       & 163.252 +10.378  & 18.3  & Sph 	 &  -     &   10.8   &  mem    & 479 &         \\
{\bf 										       	     	    
Dw1053+1518}   & 163.443 +15.311  & 21.3  & Tr           &  -     &   10.8   &  mem    & 582 &        \\
dw1055+11      & 163.930 +11.969  & 17.9  & Sph 	 &  -	  &   10.8   &  mem    & 380 &         \\
LSBC D640-12   & 163.981 +12.338  & 18.4  & Tr  	 & 699    &   10.8   &  mem    & 374 &         \\
LSBC D640-13   & 164.059 +12.009  & 16.9  & Irr 	 & 840    &   10.8   &  mem    & 401 &         \\
LSBC D640-14   & 164.543 +11.998  & 18.5  & Sph 	 &  -	  &   10.8   &  mem    & 487 &         \\
AGC205278      & 164.717 +14.130  & 17.3  & Irr 	 & 548    &   11.80  &  TF     & 585 &  (8)     \\
dw1059+11      & 164.962 +11.427  & 19.2  & Tr 	         &  -	  &   10.8   &  mem    & 593 &         \\
NGC3489	       & 165.077 +13.901  & 11.1  & S0    	 & 556    &   12.08  &  SBF    & 624 &  (7)    \\
{\bf 											     		
Dw1100+1125}   & 165.131 +11.431  & 21.5  & Tr           &  -     &   10.8   &  mem    & 622 &        \\
LeG33          & 165.188 +14.172  & 18.6  & Tr  	 &  -	  &   10.8   &  mem    & 596 &         \\
LSBC D640-08   & 165.216 +13.881  & 17.0  & Sph 	 &  -	  &   10.8   &  mem    & 646 &         \\
dw1101+11      & 165.344 +11.753  & 19.7  & Irr  	 &  -	  &   10.8   &  mem    & 641 &         \\
CGCG 066-109   & 166.109 +11.756  & 16.2  & Irr 	 & 629    &   10.30  &  TF     & 778 &  (8)     \\
{\bf 											     		
Dw1106+1250}   & 166.541 +12.845  & 18.8  & Tr           &  -     &   10.8   &  mem    & 843 &        \\

\hline	
\multicolumn{8}{l}{\small{\textbf{Note.} (1) \citet{cohen2018}; (2) \citet{carlsten2022}; (3) \citet{shaya2017}; (4) \citet{ferrarese2000}; }}\\
\multicolumn{8}{l}{\small{(5) LVGDB; (6) \citet{lee2016}; (7) \citet{tonry2001}; (8) \citet{karachentsev2015}}}
\end{longtable}

\begin{table*}                                                                             \begin{center} 
\caption{Background galaxies with $V_{\rm LG} < 2000$~km~s$^{-1}$ in the same area outlined in Fig. 3} \label{table2} 
\begin{tabular}{lccrcllc}  \hline\hline
Name        &RA (J2000.0 DEC)    &  B,   &   $V_{\rm LG}$,&       $D, $&       Method   &          $W_{50}$  & Refs \\
            &                    & mag & km s$^{-1}$   & Mpc     &            &            km s$^{-1}$   & \\ \hline

AGC202244   &   157.920 +13.835  & 16.5 &  1141      &        34.3  &     TF   &      102   &  (1)   \\
AGC202016   &   158.330 +10.190  & 19.1 &  1270      &         -    &          &       32   &        \\
AGC205161   &   158.524 +15.780  & 17.9 &  1081      &         -    &          &      114   &        \\
NGC 3279    &   158.678 +11.197  & 13.9 &  1236      &        32.2  &     TF   &      347   &  (2)   \\
AGC202248   &   158.734 +11.492  & 17.5 &  1020      &        28.2  &     TF   &       62   &  (3)   \\
LeG 03      &   158.953 +08.481  & 17.3 &   987      &        21.9  &     TF   &       70   &  (3)   \\
FGC125a     &   159.370 +12.396  & 17.4 &  1178      &        25.0  &     TF   &       59   &  (1)   \\
CGCG065-074 &   159.533 +10.381  & 14.4 &  1013      &        25.1  &     TF   &      178   &  (1)   \\
{\bf AGC203080}   &   160.421 +13.825  & 17.5 &  1127      &        23.1  &     mem  &        -   &        \\ 
{\bf NGC 3338}    &   160.531 +13.747  & 11.4 &  1155      &        23.1  &     TF   &      339   &  (2)   \\
{\bf AGC203082}   &   160.610 +13.957  & 17.8 &  1137      &        17.5  &     TF   &       41   &  (1)   \\
{\bf UGC 5832}    &   160.702 +13.460  & 14.3 &  1081      &        18.6  &     TF   &      102   &  (1)   \\ 
{\bf AGC200543}   &   160.773 +13.510  & 16.2 &  1117      &        18.1  &     TF   &       70   &  (1)   \\
{\bf NGC 3346}    &   160.912 +14.872  & 12.6 &  1135      &        18.5  &     TF   &      162   &  (4)   \\
AGC205270   &   161.291 +15.450  & 17.1 &  1083      &         -    &          &       51   &        \\
NGC 3389    &   162.116 +12.533  & 12.5 &  1159      &        25.2  &     SN   &      266   &  (5)   \\
Mrk 1263    &   162.237 +12.195  & 15.7 &  1171      &        34.2  &     TF   &      125   &  (1)   \\
AGC200600   &   162.249 +10.835  & 16.2 &  1782      &        38.4  &     TF   &      120   &  (2)   \\
AGC200603   &   162.322 +12.422  & 15.7 &  1234      &        15.6  &     TF   &       68   &  (3)   \\
PGC032376   &   162.327 +12.378  & 18.0 &  1200      &          -   &          &       25   &        \\
AGC202253   &   162.361 +12.258  & 17.5 &  1188      &          -   &          &        -   &        \\ 
AGC205197   &   162.428 +13.828  & 19.0 &  1190      &        22.7  &     bTF  &       42   &  (3)  \\ 
AGC205198   &   162.507 +13.785  & 17.2 &  1168      &        19.1  &     TF   &       53   &  (1)   \\
UGC 5948    &   162.658 +15.763  & 16.6 &   987      &        26.3  &     bTF  &      106   &  (1)  \\
UGC 6014    &   163.428 +09.728  & 15.0 &   965      &        17.0  &     TF   &       94   &  (1)   \\
SDSSJ10540  &   163.500 +09.831  & 17.5 &  1023      &        17.0  &     mem  &        -   &        \\
AGC202033   &   163.765 +14.093  & 18.9 &  1968      &          -   &          &       40   &        \\
AGC202260   &   164.409 +13.979  & 17.5 &  1078      &        28.8  &     bTF  &       92   &  (1)  \\
NGC 3485    &   165.010 +14.842  & 12.7 &  1301      &         -    &          &      135   &        \\ 
AGC202040   &   165.757 +08.048  & 18.1 &  1194      &        37.6  &     bTF  &       96   &  (1)  \\
AGC215256   &   165.860 +16.016  & 16.9 &  1102      &        37.5  &     TF   &      105   &  (1)   \\
AGC219117   &   165.945 +08.572  & 18.7 &  1575      &         -    &          &       68   &        \\
NGC 3524    &   166.634 +11.385  & 13.4 &  1216      &         -    &          &      415   &        \\
AGC215262   &   166.648 +12.231  & 18.2 &  1461      &        36.0  &     bTF  &       63   &  (3)   \\
UGC 6169    &   166.764 +12.060  & 14.6 &  1405      &        34.6  &     TF   &      241   &  (1)   \\
CGCG 067-014&   167.347 +10.834  & 15.3 &  1404      &        21.1  &     mem  &       66   &       \\
NGC 3547    &   167.483 +10.721  & 13.2 &  1438      &        21.1  &     TF   &      204   &  (6)   \\
  \hline
\multicolumn{8}{l}{\small{\textbf{Note.} (1) \citet{karachentsev2015}; (2) \citet{springob2009}; (3) this work;}}\\
\multicolumn{8}{l}{\small{(4) \citet{bottinelli1984}; (5) \citet{pejcha2015}; (6) \citet{tully2013}}}
  
  \end{tabular}
  \end{center}
  \end{table*}

 \begin{center}
\begin{large}
  \begin{table*}
   \caption{Suggested members of the ${\rm M}\,66$ group.} 
   \label{table3} 
  \begin{tabular}{lcclcrllc} \hline\hline  
      Name       &   RA(J2000.0)DEC&  B  & Type & $V_{\rm LG}$& $D$ & Method &   $R_p$ & Refs  \\                                    
                 &      deg       &  mag &      &  km s$^{-1}$& Mpc &   &  kpc & \\  \hline 
 
 Dw1111+1338	& 167.896 +13.644 &20.5 & Sph   &  	  &  11.12   &    mem &  430 &        \\
 AGC215282      & 168.605 +15.534 &16.5 & Im    & 731     &  11.12   &    mem &  565 &        \\
 NGC3593  	& 168.654 +12.818 &11.9 & S0a   & 492	  &  10.80   &    TF  &  269 &(1)     \\
 AGC202256	& 168.688 +12.648 &17.5 & Irr   & 490	  &  11.00   &    TF  &  269 &(1)     \\
 dw1114+1307	& 168.713 +13.132 &20.1 & Sph   &  	  &  11.12   &    mem &  256 &        \\
 IC2684         & 169.254 +13.099 &16.2 & Tr    & 451	  &  11.12   &    mem &  154 &        \\
 dw1117+1250    & 169.435 +12.837 &20.6 & Sph   &	  &  11.12   &    mem &  123 &        \\
 dw1118+1330	& 169.566 +13.514 &20.6 & Sph   &	  &  11.12   &    mem &  137 &        \\
 dw1118+1233	& 169.676 +12.561 &17.5 & Sph   &	  &   9.67   &    SBF &  111 &(2)    \\
 dw1118+1348	& 169.723 +13.805 &19.2 & Sph   &	  &   9.86   &    SBF &  170 &(2)   \\
 NGC3623 (M65)	& 169.733 +13.092 &10.3 &  Sa   &666	  &  12.19   &    TF  &   66 &(3)    \\
 AGC215286	& 169.803 +14.327& 16.8 &  Tr   &867	  &  11.70   &    TF  &  263 &(4)    \\
 AGC202257	& 169.810 +11.952& 16.5 &  Irr  &719	  &  11.70   &    bTF &  206 &(4)   \\
 AGC215354	& 169.816 +14.290 &17.4 &  BCD  &659	  &  11.12   &    mem &  256 &       \\
 dw1119+1404	& 169.840 +14.076 &16.9 & Sph   &354	  &   9.48   &    SBF &  214 &(2)   \\
 Dw1119+1011	& 169.878 +10.199 &18.1 &  BCD  &	  &  11.12   &    mem &  544 &       \\
 NGC3627 (M66)	& 170.062 +12.992 &9.7  & Sb    &579	  &  11.12   &    TRGB&   0  &(5)   \\
 dw1120+1332	& 170.067 +13.549 &18.3 & Im    &         &  11.43   &    SBF &  108 &(2)   \\
 NGC3628	& 170.071 +13.590 &10.3 &  Sb   &710	  &  10.52   &    TF  &  116 &(6)    \\ 
 dw1120+1337	& 170.255 +13.607 &14.9 &  Tr   &681	  &  11.12   &    mem &  125 &       \\
 dw1121+1411	& 170.372 +14.197 &20.1 &  Sph  &	  &  11.12   &    mem &  240 &       \\
 dw1121+1326	& 170.404 +13.447 &19.9 &  Irr  &	  &   8.69   &    SBF &  109 &(2)    \\
 AGC213436	& 170.600 +12.980 &16.9 &  dE   &491	  &  11.66   &    SBF &  101 &(2)   \\
 IC2782	        & 170.731 +13.441 &15.0 &  dE   &719	  &  10.46   &    SBF &  153 &(2)   \\
 AGC215414	& 170.806 +13.714 &17.8 &   Irr & 746	  &  11.12   &    mem &  198 &       \\   
 IC2787    	& 170.830 +13.630 &15.3 &  BCD  &576	  &  11.06   &    SBF &  190 &(2)   \\
 IC2791	        & 170.907 +12.896 &17.1 &  Im   &530	  &   9.95   &    SBF &  161 &(2)   \\
 dw1123+1346	& 170.985 +13.778 &20.1 &  Sph  &	  &  11.12   &    mem &  232 &       \\
 AGC215415	& 171.144 +12.674 &18.6 &  Tr   & 868	  &  11.12   &    mem &  213 &       \\
 dw1125+1316	& 171.479 +13.269 &17.3 &  dE   &   	  &  11.12   &    mem &  273 &       \\
 Dw1127+1346	& 171.804 +13.781 &20.1 &  Sph  &	  &  11.12   &    mem &  357 &       \\
 AGC213091	& 172.394 +10.810 &18.5 &  Sph  & 600	  &  11.12   &    mem &  612 &       \\
 KKH68          & 172.722 +14.145 &16.6 &   Irr & 753	  &   8.50   &    TF  &  552 &(1)    \\
 AGC215248      & 173.462 +14.054 &17.9 &  BCD  & 808     &   9.90   &    mem &  676 &        \\   
 KKH69	        & 173.721 +11.021 &16.6 &   Irr & 741	  &   7.40   &    TF  &  791 &(1)    \\ \hline
 \multicolumn{9}{l}{\small{\textbf{Note.} (1) \citet{karachentsev2013}; (2) \citet{carlsten2022}; (3) \citet{tully2013}; (4) LVGDB; }}\\
\multicolumn{9}{l}{\small{(5) \citet{hoyt2019}; (6) \citet{sorce2014}}}
\end{tabular}
\end{table*}

\begin{table}
   \caption{ Basic parameters of ${\rm M}\,105$ and ${\rm M}\,66$ group of galaxies.} 
   \label{table4} 
  \begin{tabular}{lcc} \hline\hline 
Parameter                                   & M 105 group      &     M 66 group \\
\hline
Number of known members                               &      83      &          35\\
Average radial velocity, km s$^{-1}$                  &  654$\pm$24  &      646$\pm$9\\
Radial velocity dspersion, km s$^{-1}$                &     136      &         135\\
Distance, Mpc                                         & 10.8$\pm$0.5 &     11.1$\pm$0.5\\
Average projected radius, $\langle R_p\rangle$, kpc   &     318      &         270 \\
Total $K$-band luminosity, $10^{11}~L_{\odot}$        &     3.2      &         3.8 \\
$M_{\rm orb}$, $10^{12}~M_{\odot}$                    &5.76$\pm$1.32 &    5.94$\pm$1.50\\
$R_{\rm vir}$, kpc                                    &     385      &         390\\  
Total mass-to-luminosity ratio, $M_{\odot}/L_{\odot}$ &17.8$\pm$4.1  &    15.6$\pm$3.9\\
\hline

\end{tabular}
\end{table}
  
\end{large}
\end{center}

  \end{document}